\documentclass[]{spie}  

\usepackage{amsmath,amsfonts,amssymb}
\usepackage{graphicx}
\usepackage[colorlinks=true, allcolors=blue]{hyperref}

\title{Meta-QSM: An Image-Resolution-Arbitrary Network for QSM Reconstruction}

\author[ab]{Juan Liu}
\author[abc]{Kevin M. Koch}
\affil[a]{Center for Imaging Research, Medical College of Wisconsin, Milwaukee, WI, USA}
\affil[b]{Joint Department of Biomedical Engineering, Marquette University and Medical College of Wisconsin, Milwaukee, WI, USA}
\affil[c]{Radiology, Medical College of Wisconsin, Milwaukee, WI, USA}

\authorinfo{Further author information: (Send correspondence to Kevin M. Koch)\\Kevin M. Koch.: E-mail: kmkoch@mcw.edu, Telephone: 1 414 955 4034}

\begin{document} 
\maketitle

\begin{abstract}
Quantitative Susceptibility Mapping (QSM) can estimate the underlying tissue magnetic susceptibility and reveal pathology. Current deep-learning-based approaches to solve the QSM inverse problem are restricted on fixed image resolution. They trained a specific model for each image resolution which is inefficient in computing. In this work, we proposed a novel method called Meta-QSM to firstly solve QSM reconstruction of arbitrary image resolution with a single model. In Meta-QSM, weight prediction was used to predict the weights of kernels by taking the image resolution as input. The proposed method was evaluated on synthetic data and clinical data with comparison to existing QSM reconstruction methods. The experimental results showed the Meta-QSM can effectively reconstruct susceptibility maps with different image resolution using one neural network training.

\end{abstract}


\section{INTRODUCTION}
\label{sec:intro}  

Quantitative Susceptibility Mapping (QSM) can estimate the underlying tissue magnetic susceptibility to provide a novel image contrast and reveal pathology. QSM requires addressing a challenging post-processing problem: filtering of image phase estimates and inversion of the phase to susceptibility relationship. Both steps require solving ill-posed inverse computational problems, causing a wide variety of quantification errors, robustness limitations, and artifacts. 

To date, all QSM methods rely on a dipolar convolution that relates source susceptibility to induced Larmor frequency offsets ~\cite{salomir2003fast, marques2005application}. While the forward relationship of this model (source to field) can be efficiently computed using Fast-Fourier-Transforms (FFT), a k-space singularity in the applied convolution filter results in an ill-conditioned relationship in the inverse model (field to source). Acquiring multiple orientations data to the magnetic field can sufficiently improve the conditioning of the inversion algorithm, serving as the empirical gold-standard for \emph{in vivo} QSM assessment~\cite{liu2009calculation}. Single-orientation susceptibility maps computation is more challenging. Thresholding of the convolution operator~\cite{shmueli2009magnetic, wharton2010susceptibility, haacke2010susceptibility} or use of more sophisticated regularization methods \cite{de2008quantitative, de2010quantitative, liu2011morphology, bilgic2014fast} are applied. However, these methods suffer from long computation time, regularization parameter tuning, conspicuity loss of fine details, and streaking artifacts, causing technical implementation challenges of QSM clinical translation and practice.

Current deep-learning-based methods such as QSMnet\cite{yoon2018quantitative}, DeepQSM\cite{bollmann2019deepqsm} have shown improved QSM reconstruction results compared with conventional methods. However, they trained a specific model for each image resolution, which requires train multiple models for different image resolution. If we train a specific model for each image resolution, it is impossible to store all these models and it is inefficient in computing. Thus, a method to solve QSM reconstruction of arbitrary image resolution is important for putting the deep-learning-based QSM approaches into practical use.  

To solve these drawbacks, we proposed a method called Meta-QSM. It adopts the weight prediction method used in the meta-learning, to dynamically predict weights of the filters for each image resolution. Quantitative evaluation was conducted using synthetic test data with comparison to conventional QSM reconstruction methods. Besides, qualitative assessment was performed using two QSM datasets acquired with different image resolution. 

\section{METHODS}

\subsection{Training Data}

We used one QSM dataset and data augmentation techniques to generate whole synthetic training data. The COSMOS dataset from 2016 ISMRM QSM challenge \cite{langkammer2018quantitative} was used, which was acquired with a fast 3-dimensional gradient-echo scans, 12 different head orientations, 1.06mm isotropic voxels on a 3T scanner. The COSMOS data was resampled to image resolution 0.75x0.75x3.0 mm$^3$. Elastic distortions are applied to geometrically transform the susceptibility map \cite{simard2003best}. In addition, randomly sized geometric shapes, such as ellipsoids, spheres, cuboids, and cylinders with random susceptibility values are randomly placed on the augmented susceptibility maps to mimic the bleeding and calcifications, in order to increase the training data variability. Image contrast changes are applied to susceptibility maps as well to increases the training data variability in different image contrast. The image resolution of generated QSM data was randomly set to (0.5-1.0, 0.5-1.0, 2.0-4.0)mm$^3$. The local fields are calculated using the well-defined forward dipole convolution relationship. The local fields and QSM images are cropped to image size 128x128x64 for training.

\begin{figure}[ht]
\begin{center}
\includegraphics[width=14cm]{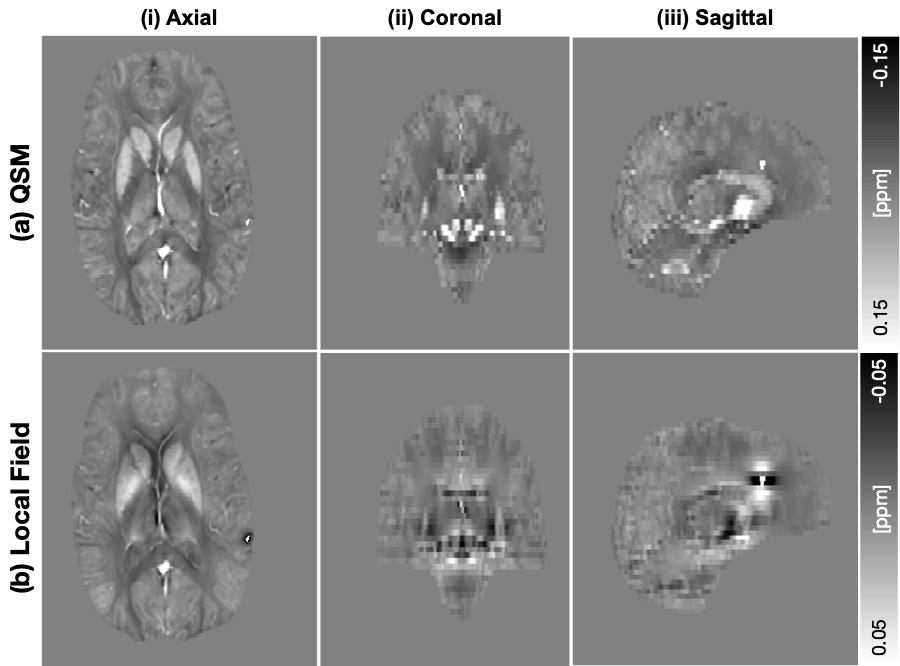}
\caption{One example of training data. 
\label{fig_trainingdata}}
\end{center}
\end{figure}

\subsection{Neural Network Architecture}

\begin{figure}[ht]
\begin{center}
\includegraphics[width=14cm]{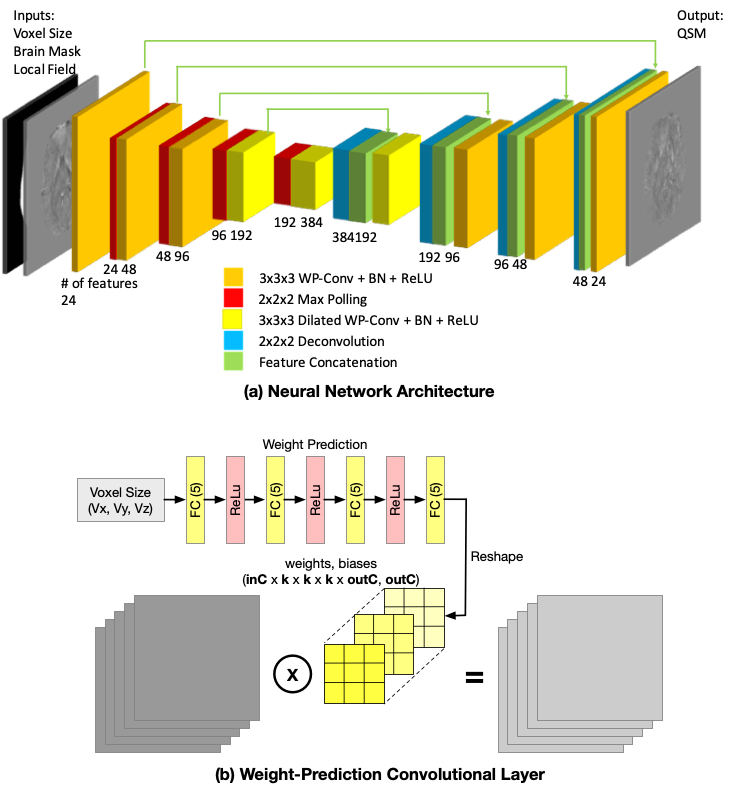}
\caption{Network structure of Meta-QSM. A 3D convolutional neural network was designed with 6 WP-Conv (kernel size 3x3x3, dilated rate 1x1x1), 3 WP-Conv (kernel size 3x3x3, dilated rate 2x2x2), 4 max pooling layer (pool size 2x2x2, stride size 2x2x2), 4 deconvolutional layers (stride size 2x2x2), 9 batch normalization layers, 4 feature concatenations, and 1 WP-Conv (kernel size 3x3x3, linear activation). 
\label{fig_metaQSM}}
\end{center}
\end{figure}

A 3D encoder-decoder convolutional neural network, based on a modified version of an established U-Net architecture \cite{ronneberger2015u}, was trained to perform pixel-wise susceptibility estimation. All vanilla convolution layers were replaced by weight-predict convolutional layers (WP-Conv), as shown in Fig.\ref{fig_metaQSM}. For WP-Conv, it consists of several fully connected layers and several activation layers. WP-Conv takes the image resolution as input, and the learn resolution specific weights and bias with the shape (inC, k, k, k, outC) and (outC) respectively. Here the inC is the number of channels of the input feature map. k is the kernel size, which is 3 in the paper. The outC is the number of channels of the output feature map. 

\subsection{Training Details}

L1 loss between the susceptibility map and label was utilized as a loss function for training. L2 regularization was added in all WP-Conv layers with regularization strength 0.01. 30,000 synthetic data were spilt into training dataset and validation dataset with ratio 9:1. The RMSprop optimizer was used in the deep learning training. The initial learning rate was set as 0.0001, with exponential decay at every 200 steps. Two NVIDIA tesla k40 graphics processing units (GPUs) were used for training with batch size 4. The neural network was trained and evaluated using Keras with Tensorflow as backend.

\section{PERFORMANCE EVALUATION}
\label{sec:eval}

\subsection{Synthetic Data} 

100 synthetic data sets with matrix size 256x256x64 generated in similar fashion to the training data were used to evaluate the performance of Meta-QSM with comparison with Truncated K-Space Division (TKD) inversion \cite{shmueli2009magnetic}, iLSQR\cite{li2015method}, and Morphology Enabled Dipole Inversion (MEDI) \cite{liu2012morphology}. The results were evaluated against the ground-truth using root mean squared error (RMSE), high-frequency error norm (HFEN), and structural similarity (SSIM) index.
  
\subsection{Comparison with Mesa-QSM}
We trained two neural networks, denoted as Mesa-QSM, for image resolution 0.5x0.5x2.0mm$^3$ and  0.7639x0.7639x3.0mm$^3$. Mesa-QSM replaces the WP-Conv with vanilla convolution layers in Meta-QSM. The neural networks was trained using the same hyperparameters.  

After training, 100 synthetic data generated with image resolution 0.5x0.5x2.0mm$^3$ and  0.7639x0.7639x3.0mm$^3$ in in similar fashion to the training data were used to evaluate the performance of Meta-QSM and Mesa-QSM. The results were evaluated against the ground-truth using RMSE, HFEN, and SSIM index.

\subsection{3T Research Data} 

MR imaging for a large local sports concussion study approved by the local institutional human research review board was performed on a 3T MRI scanner (GE Healthcare MR750) using a 32ch MRI head receive array using a commercially available susceptibility-weighted software application (SWAN, GE Healthcare). The data acquisition parameters were as follows: in-plane data matrix - 320x256, field of view - 24 cm, voxel size - 0.5x0.5x2.0 mm$^3$, echo spacing - 7 ms, 4 echo times - [10.4, 17.4, 24.4, 31.4] ms, repetition time - 58.6 ms, autocalibrated parallel imaging factors - 3x1, acquisition time - 4 min.

Complex multi-echo images were reconstructed from raw k-space data using GE Orchestra. The brain masks were obtained using the SPM tool~\cite{brett2002region}. After background field removal using the Regularization-enabled Sophisticated Harmonic Artifact Reduction on Phase data (RESHARP) \cite{sun2014background} method, susceptibility inversion was performed using the TKD, iLSQR, MEDI, Mesa-QSM, and Meta-QSM. 

\subsection{3T Clinical SWI Data} 

The clinical QSM data were acquired using susceptibility weighted angiography (SWAN, GE Healthcare), at a 3T MRI scanner (GE Healthcare MR750) with data acquisition parameters: in-plane data acquisition matrix 288x224, field of view 22 cm, slice thickness 3 mm, autocalibrated parallel imaging factors 1x2 or 1x3, number of slices 46-54, first echo time 12.6 ms, echo spacing 4.1 ms, number of echoes 7, flip angle 15$^o$, repetition time 39.7 ms, total scan time about 2 minutes.  

The SWI images were processed by vendor reconstruction algorithms. The raw k-space data were saved for offline QSM processing. Multi-echo real and imaginary data were reconstructed from k-space data, with reconstruction matrix size 288x288, voxel size 0.76x0.76x3.0 mm$^3$. The field map was obtained by the fitting of multi-echo phases. Brain masks were obtained using the SPM brain extraction tool. RESHARP were applied to remove the background field, with spherical kernel radius set as 6mm. Susceptibility inversion was performed using the TKD, iLSQR, MEDI, Mesa-QSM, and Meta-QSM.

In the above, for RESHARP, spherical kernel radius was set as 6mm to trade off the background removal performance and brain erosion. For TKD, the threshold was set to 0.20; for MEDI, the regularization factor was set to 1000. MEDI toolbox and STI toolbox publicly provided by their respective authors were used to calculate the QSM images use for comparison analyses in this report. 

\section{RESULTS}
\label{sec:results}

\subsection{Synthetic Data} 

\begin{table}[ht]
\caption{Quantitative comparison Meta-QSM with conventional QSM reconstruction methods} 
\label{table_sim_data}
\begin{center}       
\begin{tabular}{|l|l|l|l|}
\hline
\rule[-1ex]{0pt}{3.5ex} & TKD  & iLSQR  & Meta-QSM  \\
\hline
\rule[-1ex]{0pt}{3.5ex}  RMSE($\%$) & $32.90\pm1.55$ & $52.94\pm3.54$ & $21.96\pm3.84$  \\
\hline
\rule[-1ex]{0pt}{3.5ex}  HFEN($\%$) & $33.07\pm1.65$ & $51.11\pm3.77$ & $16.75\pm3.84$  \\
\hline
\rule[-1ex]{0pt}{3.5ex}  SSIM(0-1) & $0.960\pm0.013$ & $0.941\pm0.023$ & $0.977\pm0.013$  \\
\hline
\end{tabular}
\end{center}
\end{table}

Table.\ref{table_sim_data} illustrates error metrics using TKD, iLSQR, Meta-QSM from 100 synthetic data. The proposed method achieved the best score in RMSE, HFEN, and SSIM. 

\subsection{Comparison with Mesa-QSM}

\begin{table}[ht]
\caption{Quantitative comparison Meta-QSM with Mesa-QSM on image resolution 0.5x0.5x2.0mm$^3$.} 
\label{table_mesa_cmp_p5p5x2}
\begin{center}       
\begin{tabular}{|l|l|l|l|l|l|}
\hline
\rule[-1ex]{0pt}{3.5ex}   & Mesa-QSM & Meta-QSM   \\
\hline
\rule[-1ex]{0pt}{3.5ex}  RMSE($\%$)  & $25.13\pm1.75$ & $28.33\pm1.70$\\
\hline
\rule[-1ex]{0pt}{3.5ex}  HFEN($\%$)  & $17.07\pm2.27$ & $22.10\pm2.11$ \\
\hline
\rule[-1ex]{0pt}{3.5ex}  SSIM(0-1)   & $0.968\pm0.012$ & $0.962\pm0.014$ \\
\hline
\end{tabular}
\end{center}
\end{table}

\begin{table}[ht]
\caption{Quantitative comparison Meta-QSM with Mesa-QSM on image resolution 0.76x0.76x3.0mm$^3$.} 
\label{table_mesa_cmp_p7p7x3}
\begin{center}       
\begin{tabular}{|l|l|l|l|l|l|}
\hline
\rule[-1ex]{0pt}{3.5ex}  & Mesa-QSM & Meta-QSM  \\
\hline
\rule[-1ex]{0pt}{3.5ex}  RMSE($\%$) & $25.34\pm1.74$ & $18.87\pm2.00$  \\
\hline
\rule[-1ex]{0pt}{3.5ex}  HFEN($\%$) & $17.43\pm2.38$ & $14.04\pm2.05$  \\
\hline
\rule[-1ex]{0pt}{3.5ex}  SSIM(0-1) & $0.968\pm0.012$ & $0.984\pm0.007$ \\
\hline
\end{tabular}
\end{center}
\end{table}

Table.\ref{table_mesa_cmp_p5p5x2} and Table.\ref{table_mesa_cmp_p7p7x3} shows the quantitative metrics of comparison Meta-QSM with two Mesa-QSM. For image resolution 0.76x0.76x3.0mm$^3$, Meta-QSM achieved better metric scores than Mesa-QSM. For image resolution 0.5x0.5x2.0mm$^3$, Mesa-QSM achieved slightly better metric scores than Meta-QSM. 

\subsection{3T Research Data} 

\begin{figure}[ht]
\begin{center}
\includegraphics[width=14cm]{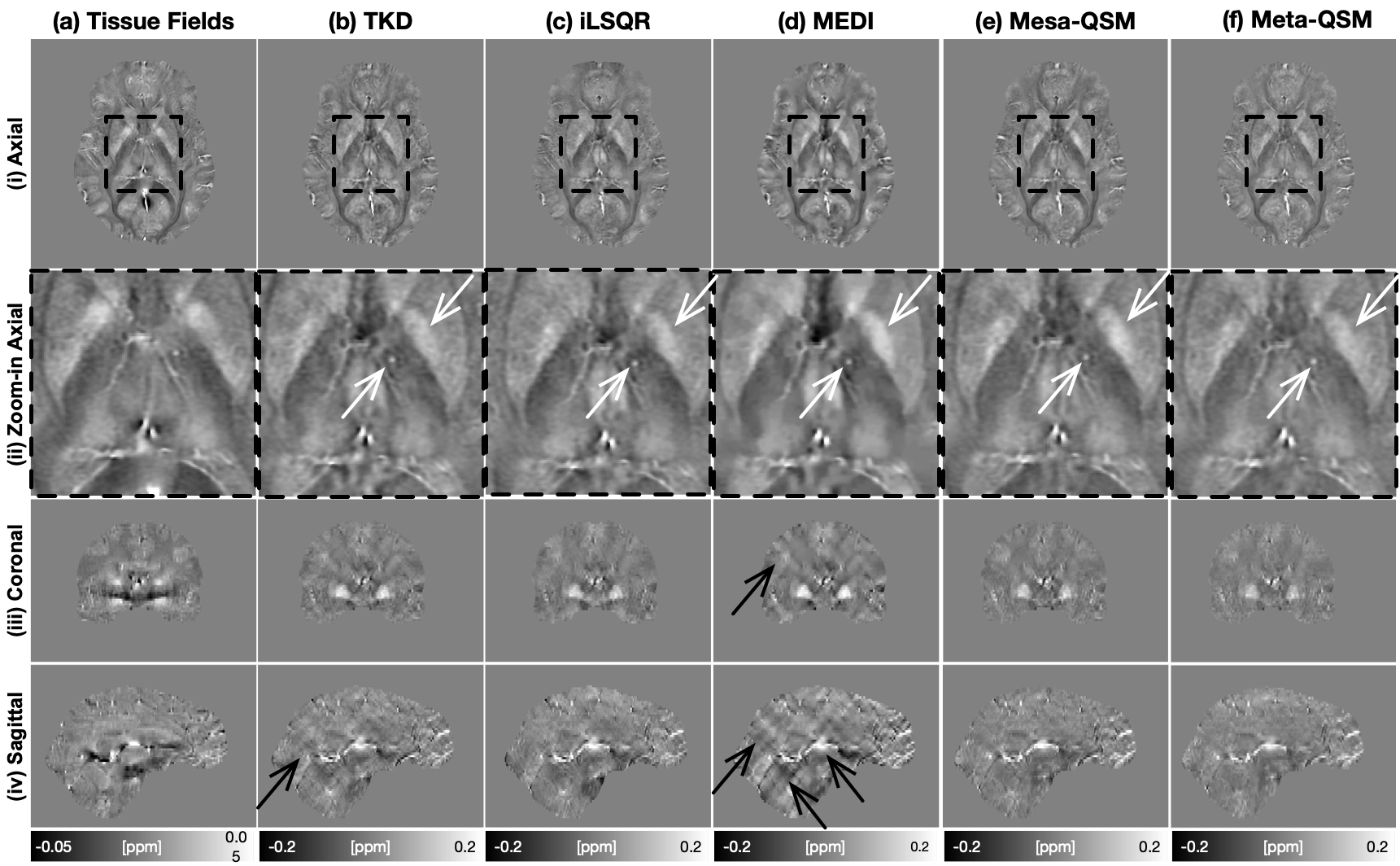}
\caption{Tissue fields (a) and QSM images (b-f) in a 3T MRI data with image resolution 0.5x0.5x2.0mm$^3$. 
\label{fig_concussion}}
\end{center}
\end{figure}

Fig.\ref{fig_concussion} shows QSM images reconstructed by TKD, iLSQR, MEDI, Mesa-QSM, and Meta-QSM. Image blurring showed in  TKD and MEDI results. Mesa-QSM and Meta-QSM can produce images with high image sharpness. Streaking artifacts is clearly observable in TKD and MEDI results. Meta-QSM showed a slight underestimated susceptibility values.

\subsection{3T Clinical SWI Data} 

\begin{figure}[ht]
\begin{center}
\includegraphics[width=16cm]{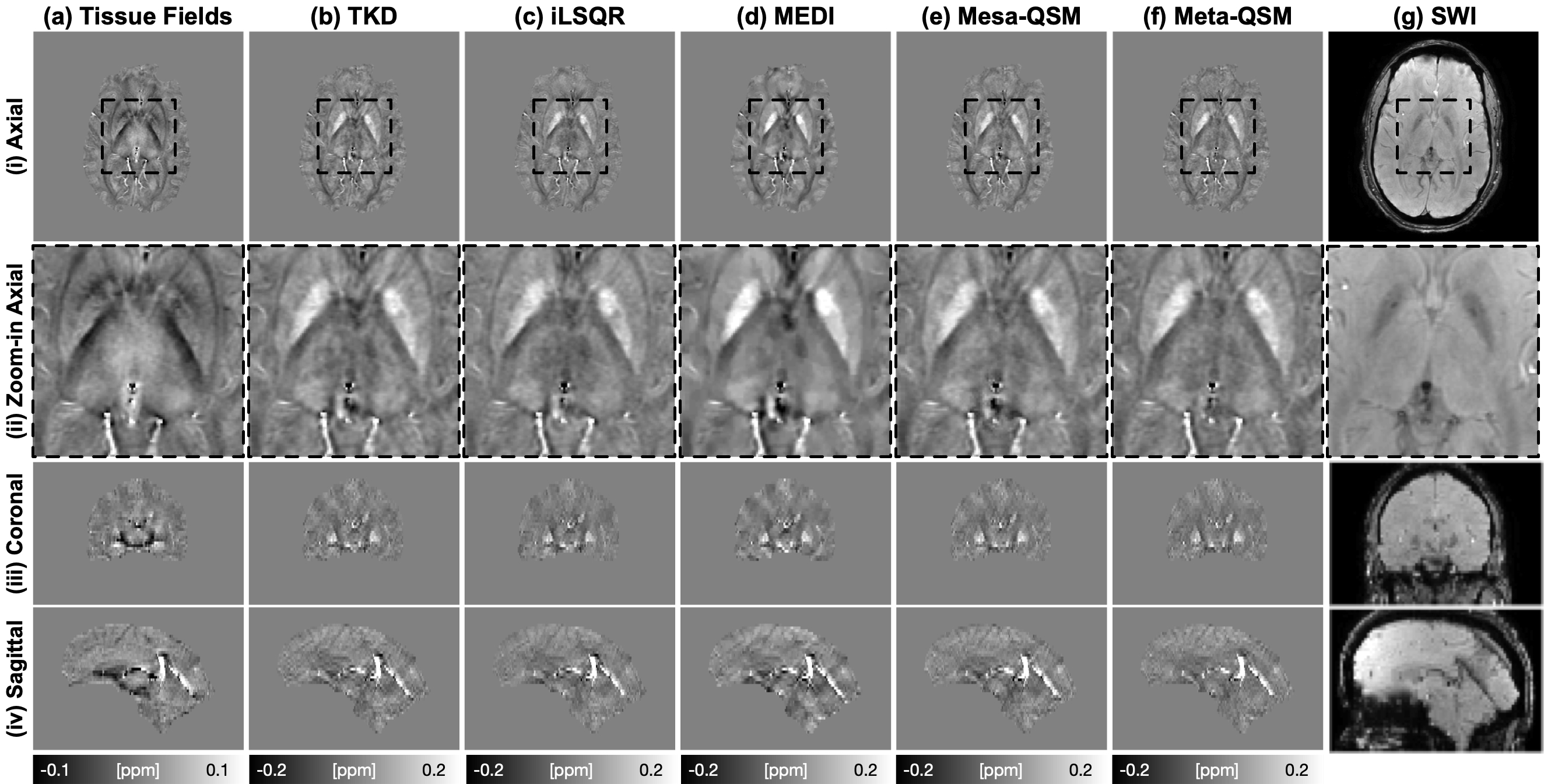}
\caption{Tissue fields and QSM images (b-f), SWI images from a 26-year-old patient with dizziness and cognitive impairment after stroke.
\label{fig_clinical1}}
\end{center}
\end{figure}

\begin{figure}[ht]
\begin{center}
\includegraphics[width=16cm]{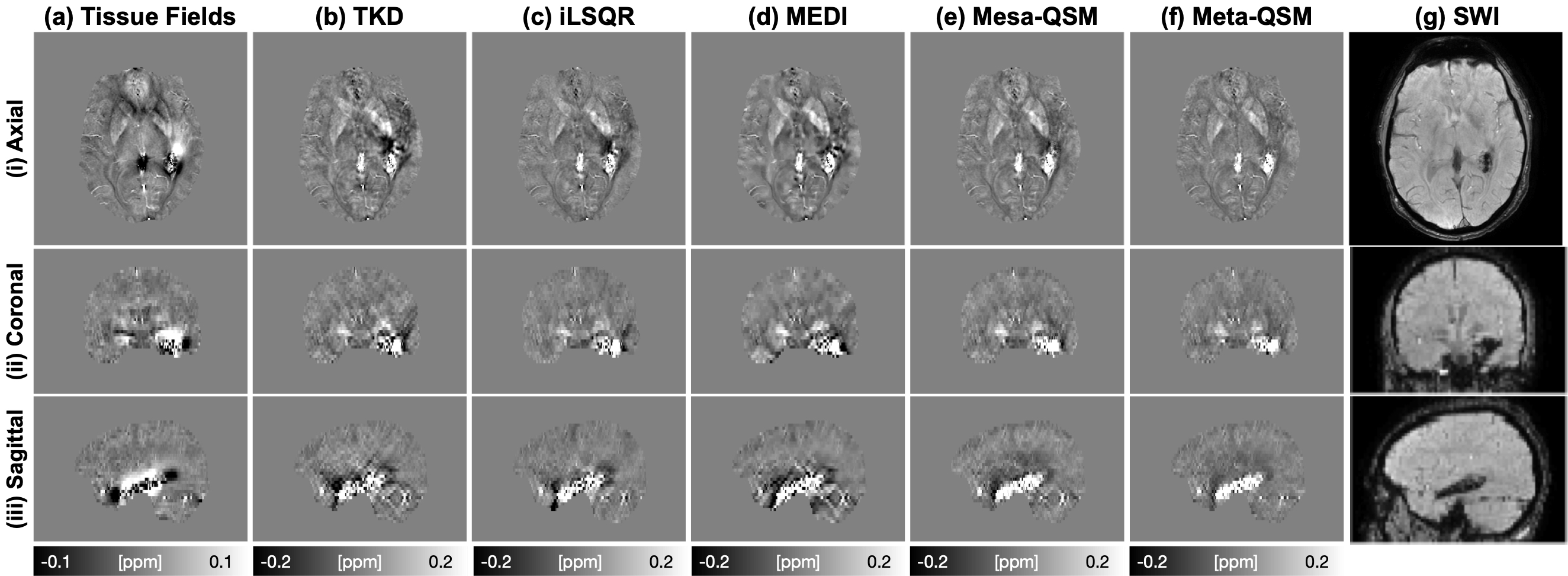}
\caption{Tissue fields and QSM images (b-f), SWI images from a 28-year-old patient with left mesial temporal lesion and neurofibromatosis type 1.
\label{fig_clinical2}}
\end{center}
\end{figure}

\begin{figure}[ht]
\begin{center}
\includegraphics[width=16cm]{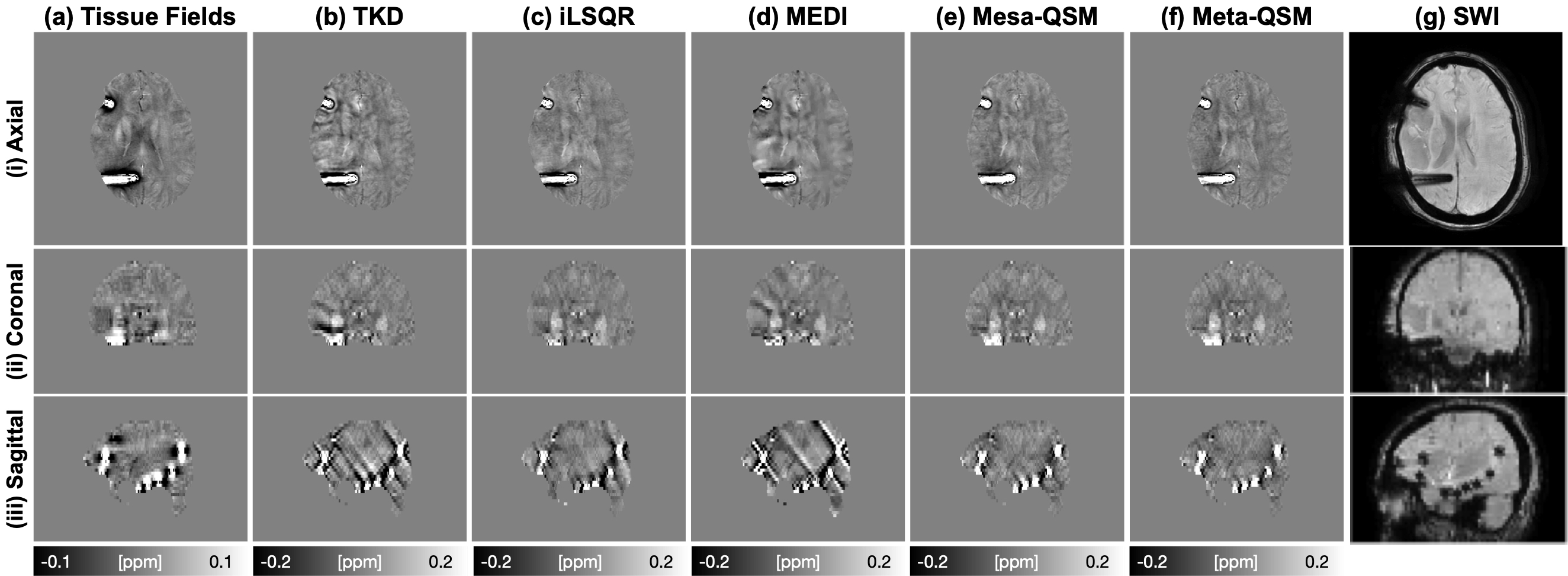}
\caption{Tissue fields and QSM images (b-f), SWI images from a 36-year-old patient with surgical planning.
\label{fig_clinical3}}
\end{center}
\end{figure}

Fig.\ref{fig_clinical1} shows QSM images reconstructed by five methods and SWI images on a 26-year-old patient with dizziness and cognitive impairment after stroke. Meta-QSM can produce high quality QSM images.

Fig.\ref{fig_clinical2} shows QSM images and SWI images from a 28-year-old patient with left mesial temporal lesion and neurofibromatosis type 1. TKD, iLSQR, and MEDI results suffer from image blurring and shading artifacts around hemorrhage. Meta-QSM results show the best image quality.

Fig.\ref{fig_clinical3} shows QSM images and SWI images from a 36-year-old patient with surgical planning. Meta-QSM results can preserve the microstructures and suppress streaking artifacts.

\section{DISCUSSION}
\label{sec:discuss}

Quantitative and qualitative analysis within this study has demonstrated that Meta-QSM can outperform conventional QSM reconstruction methods. In qualitative comparisons with TKD, iLSQR, and MEDI, the proposed method can outperform based on quantitative metrics evaluation. Qualitatively, Meta-QSM can not only preserve the tissue microstructure but also greatly suppress the streaking artifacts. Meta-QSM and Mesa-QSM show comparable performance in quantitative evaluation and qualitative visual assessment. For image resolution 0.5x0.5x2.0mm$^3$, Meta-QSM achieved slightly worse metric scores than Mesa-QSM, which may be due to the training data with image resolution (0.5-1.0, 0.5-1.0, 2.0-4.0)mm$^3$. 

Compared QSMnet and DeepQSM, Meta-QSM has several advantages. First, Meta-QSM applied weight prediction to solve the fixed image resolution input. Second, Meta-QSM ultilized one in-vivo QSM dataset, data augmentation techniques and well-established physical model to generate whole training data. Third, Meta-QSM training data includes simulated bleeding/calcifications, which can greatly improve performance in pathological clinical cases. 

There are several limitations of Meta-QSM. First, it is trained on synthetic data, which is due to the physical impossibility of developing a true experimental gold standard. The data distribution difference between the synthetic training data and collected test data can introduce inference errors. Second, Meta-QSM takes the local fields after background field removal to predict the susceptibility distribution. The background field removal errors can sequentially introduce susceptibility quantification errors.

\section{CONCLUSION}
\label{sec:conclusion}

We propose a novel QSM methods to solve the QSM reconstruction of arbitrary image resolution with a single model. The proposed Meta-Prediction Convolution could dynamically predict the weights of the filters. For each image resolution, the proposed Meta-Prediction Convolution generates a group of weights for the convolution layers. Thanks to the weight prediction, we can train a single model for QSM reconstruction of arbitrary image resolution. 

\bibliography{report} 
\bibliographystyle{spiebib} 

\end{document}